\documentclass[twocolumn,pra]{revtex4-1}
\usepackage{hyperref}
\usepackage{bbm,bm}
\usepackage{amsmath, amssymb}
\usepackage{xcolor}
\usepackage{graphicx,epsfig}
\usepackage{pifont}

\begin{document}
	\title{Time-reversed
	quantum trajectory analysis of micromaser correlation properties and
	fluctuation relations}
		\author{J.\ D.\ Cresser}
		\affiliation{Department of Physics and Astronomy, University of
	Glasgow, Glasgow, G12 8QQ UK}
		\affiliation{Department of Physics and Astronomy, Macquarie
	University, 2109 NSW, Australia}
		\email{james.cresser@glasgow.ac.uk}

	\begin{abstract}
		The micromaser is examined with the aim of understanding certain of
its properties based on a time-reversed quantum trajectory analysis.
The background theory of master equations derived from a repeated
interaction model perspective is briefly reviewed and extended by
taking into account the more general renewal process description of
the sequence of interactions of the system with incoming ancilla, and
results compared with other recent (and not so recent) approaches
that use this generalisation. The results are then specialised to the
micromaser, and a quantum trajectory unravelling of the micromaser
dynamics is formulated that enables time-reversed quantum
trajectories, defined according to the Crooks approach, to, first, be
shown to arise naturally in the analysis of micromaser and atomic
beam correlations, and second used in the formulation of a
fluctuation relation for the probabilities of trajectories and their
time-reversed counterparts.
	\end{abstract}

	\maketitle

\section{Introduction}

The one-atom maser, or micromaser, has been an important tool in the
experimental \cite{Meschede85,Rempe87} and theoretical
\cite{Filipowicz86,Lugiato87} study of the fundamental properties of
cavity quantum electrodynamics for over three decades. It is a
deceptively simple device: a single quantized mode of a high-$Q$
($\sim10^{10}$) cavity driven by excited two-level atoms passing
through the cavity at a rate $R$ sufficiently low that at no time is
there more than one atom in the cavity. The two atomic levels are
Rydberg states that, in free space, are very long lived, but within
the cavity, they couple strongly to the cavity microwave field. This
field is, in turn, coupled to an external thermal bath kept at a
temperature near absolute zero.

Recently, analogues of the micromaser beyond its original quantum
optical setting have been realised e.g., for a nanomechanical
resonator coupled to a superconducting single-electron transistors
\cite{Rodrigues07,You11}. Also recently, there have been a number of
theoretical developments in the theory of open quantum systems which
are of immediate relevance to the theory of micromaser. The first
such is the growing interest in repeated interaction, or
collisional models of open quantum systems, of which it can be argued
the micromaser is a very early example. A second development is in
the context of quantum thermodynamics where detailed fluctuation relations for heat exchange and entropy production have been formulated which relate the probabilities of a quantum process conditioned on the outcomes of sequences of random quantum jumps (i.e., a quantum trajectory) proceeding forward in time to its time-reversed dual quantum trajectory. 

The aim of the work to be presented here is to address these two
developments in the context of micromaser theory. The first is
motivated by the fact that recently proposed collisional models
\cite{Budini09,Vacchini13,Vacchini14}, and in particular
\cite{Vacchini16}, that make use of renewal theory to describe the
random spacing between successive collisions, can lead to somewhat
different dynamical equations, i.e., master equations, for a general
system, and hence also for the micromaser. The derivation of one such
class of equations, an extension of that done in
\cite{Cresser92b,Cresser96} is presented in Section \ref{ImpColMod},
and the origin of the differences are discussed there. The second is
motivated by the fact that certain correlation properties of the
micromaser cavity field and its associated post-interaction atomic beam,
which have been shown to be expressible in a natural way in terms of
time-reversed quantum trajectories, are shown in Section
\ref{FieldBeamCorr} to be an example of time-reversal defined in a
specific sense due to Crooks\cite{Crooks08}. This is then followed,
in Section \ref{MicroFT}, by an investigation into the micromaser
treated as a non-equilibrium thermodynamic system, leading to a
discussion of how time-reversed quantum trajectories can be defined
in this case, this in turn enabling a derivation of a detailed
fluctuation relation\cite{Crooks99} for the entropy flow between
the cavity field reservoir and the atomic beam incident on the cavity.

Conclusions and acknowledgements then close out the paper. 

\section{Background\label{Background}}

\subsection{Relation to collisional models} The micromaser
cavity-field is an open quantum system and as such its dynamics are
given by tracing over the atomic beam and thermal bath to yield a
master equation for the reduced density operator $\rho$ for the
cavity field. The original derivation of the master equation by
Filipowicz et al \cite{Filipowicz86} was recast in the
operations-effects language of open quantum systems by
\cite{Milburn87} and in terms of a quantum field description of the
atomic beam in \cite{Cresser92b,Cresser96}. But it is clear from the
structure of the model that the micromaser is an early example of a
collisional or repeated interaction model that has received
significant attention of late
\cite{Rau63,Scarani02,Ziman02,Ziman05,Ziman05a,Giovannetti05,Attal06,Pellegrini09,Giovannetti12,Giovannetti12a,Rybar12,Ciccarello13,Caruso14,Lorenzo14,Vacchini14,Kretschmer16,Vacchini16}.
The atomic beam here plays the role, in collisional models, of a
stream of elementary ancillas or `units' that interact with (i.e.,
collide with) the system $S$, then move on, making way for the next
incoming unit. Such models have proved to be a promising tool to
analyze the dynamics of quantum Markovian and non-Markovian systems
in that the master equation for the reduced system can be obtained in
many cases without any of the approximations (such as the
Born-Markov-secular approximations) usually needed for typical
microscopic derivations. As such, the micromaser expressed in the
repeated interaction picture has attracted interest from the point of view of a rigorous mathematical analysis \cite{Bruneau09}.

\subsection{The role of quantum trajectory methods} 

One feature of the model is that the atoms, after passing through the
cavity, will become entangled with the cavity field state. Moreover,
these atoms are available for detection -- the state of the atoms,
i.e., whether excited or not, can be measured by field-ionization
techniques, and consequently information on the cavity field can be
extracted from the atomic beam measurements. This inspired, very
early on, Meystre and Wright \cite{Meystre88} to carry out numerical
simulations of the micromaser dynamics based on such measurements,
which they described as `quantum trajectories', a terminology (along with the notion of `unravelling') later introduced by Alsing and Carmichael \cite{Alsing91} in the
development of the wave function Monte-Carlo or quantum trajectory
method \cite{Dum92,Dalibard92,Carmichael93}, one of the now central
theoretical tools used to analyse the properties of open quantum
systems. 

The distinction between the earlier work of Meystre and Wright and
the later developments of quantum trajectory theory is that in the
latter, the trajectories are generated by the intrinsic probabilistic
dynamics of the system whereas for the micromaser, these
probabilities are imposed externally by the arrival statistics of the
atoms in the beam. In the standard model, these arrival statistics
are taken to be Poissonian, but were later generalised using a
renewal process model in \cite{Cresser92b,Cresser96}, this leading to
a non-Markovian master equation for the cavity field. A general
quantum trajectory analysis taking into account decay of the cavity
field can be found in \cite{Cresser96,Casagrande98,Guerra01}. 

But the quantum trajectory method applied to the micromaser yields
more than simulations of cavity dynamics.  The atoms in the emergent
atomic beam will also be correlated (and indeed entangled if the mean
time between atomic arrivals is much less than the decay time of the
cavity field \cite{Phoenix07}). In the limit of a cavity field
reservoir at zero temperature, it has been shown by a quantum
trajectory based analysis that measurements of atomic beam
correlations lead directly to the cavity field intensity correlation
function (i.e., $g^{(2)}(\tau)$), \cite{Cresser94,Cresser96}, (a
result also obtained by non-trajectory methods in \cite{Herzog94}),
and by a suitable quantum interference scheme, leads to the cavity
field correlation function, $g^{(1)}(\tau)$ and hence the cavity
field spectrum \cite{Cresser06} (also obtainable via measuring the
decay of introduced coherence \cite{Casagrande03,Quang05,Vogel05}). 

These results come about because of a naturally emerging `dual'
relationship between the quantum trajectories that contribute to
$g^{(2)}(\tau)$ (and $g^{(1)}(\tau)$) and those that contribute to
the atomic beam correlation function: they are time-reversed
conjugates, and can be understood as an example of conjugate
time-reversed trajectories defined in the sense of Crooks
\cite{Crooks08}, a connection that is elaborated on further below.

The natural role of quantum trajectories in the analysis of the
micromaser is seen to play a further role when the thermodynamic
properties of the micromaser are examined.

\subsection{Thermodynamic properties of the micromaser} In its normal
mode of operation \cite{Filipowicz86}, the atoms in the incident
atomic beam are all wholly in their excited state which can be
interpreted as corresponding to an infinite negative temperature. But
if the atoms incident on the cavity are drawn from a thermodynamic
source at a finite temperature $T_{a}$, the atoms will be in a mixed
state with Boltzmann probabilities for the ground and excited states.
The micromaser then assumes the character of a thermodynamic device
operating between the two reservoirs at different temperatures, that
of the atomic source, $T_{a}$ and that of the cavity field reservoir,
$T_{c}$. The micromaser interpreted in this fashion has been
investigated from a thermodynamic perspective in a recent paper
\cite{Strasberg17} which is based explicitly on a collisional or
repeated interaction interpretation of the micromaser. This work, as
far as the micromaser is concerned, limits itself to a particular set
of thermodynamic issues, but as shown here, the  micromaser also provides a natural
setting via a quantum trajectory treatment for the analysis of
fluctuation theorems of Crooks \cite{Crooks99}.

\section{Impulsive collisional models\label{ImpColMod}}

The collisional model for open quantum systems, of which the
micromaser is an early example, involves the system of interest $S$
undergoing interaction with a succession of ancilla, and between such
interactions, the system evolves unitarily according to its own
intrinsic Hamiltonian, unless the system itself is coupled to an
external reservoir in which case a more general non-unitary evolution
will occur. In the simplest case, these ancilla are independent, and
are all prepared in the same state, and in the extreme instance, the
interaction time is sufficiently short on the time scale of evolution
of the system that this interaction is effectively impulsive, and can
be modelled as an instantaneous change in the state of the system. 

The result of the system-ancilla interaction is determined by the
detailed nature of the system, the ancilla, and their interaction,
but in general can be expressed, for a collision initiated at time $t$ as
\begin{equation}
	\rho(t+\tau_\text{int})
	=\mbox{Tr}_{a}\left[U(\tau_\text{int})
	\rho(t)\otimes\rho_aU^\dagger(\tau_\text{int})\right].
	\label{NonzeroTimeSystemJump}
\end{equation}
which, on putting $\rho_a=\sum_{n}^{}p_n|n\rangle\langle n|$
and with $\sqrt{p_n}\langle
m|U(\tau_\text{int})|n\rangle=L_{mn}(\tau_\text{int})$, these being
operators on the Hilbert space of the system, Eq.\
(\ref{NonzeroTimeSystemJump}) can be written
\begin{align}
	&\rho(t+\tau_\text{int})-\rho(t)\notag\\
	&=\sum_{mn}^{}\left(L_{mn}(\tau_\text{int})
	\rho(t)L^\dagger_{mn}(\tau_\text{int})\right.\notag\\
	&~\left.-\tfrac{1}{2}\{L^\dagger_{mn}(\tau_\text{int})
	L_{mn}(\tau_\text{int}),\rho(t)\}\right).
	\label{LindbladSystemAncilla}
\end{align}
which is of Lindblad form. We can write this as
\begin{equation}
	\rho(t+\tau_\text{int})=(1+\mathcal{F}_a
	(\tau_\text{int}))\rho(t)
	\label{JumpMap}
\end{equation}
so that the quantum map $\mathcal{F}_a (\tau_\text{int})$ represents
the change in the state of $S$ as a consequence of the interaction of
$S$ with the ancilla over the interval $\tau_\text{int}$.

In the limit of zero interaction time, which limit will also require
the interaction strength to become infinite, this can be written, with $\epsilon$ infinitesimal, as
\begin{equation}
	\rho(t+\epsilon)=(1+\mathcal{F}_a
)\rho(t-\epsilon).
\end{equation}
Between these collisions, the system will evolve freely. The system
could be assumed to be isolated or open but here we are interested in
the models in which the system is weakly coupled to a thermal
reservoir, in which case the system evolution will be described by a
Lindblad evolution
\begin{equation}
	\rho(t_n+\tau)=e^{\mathcal{L}_S\tau}\rho(t_n+\epsilon)
	\label{AncillaJump}
\end{equation}
where $\mathcal{L}_S$ is the Lindblad superoperator appropriate for
the system reservoir interaction.

Now assume that the system is initially prepared in a state
$\rho(0)$, and that a steady stream of ancilla interact with the
system in the manner described by Eq.\ (\ref{AncillaJump}), with the
first ancilla interacting at time $t_1>0$, and subsequent ancilla
arriving at times $t_2,t_3\ldots $. The state of the system at a time
$t$, $\rho_c(t)$, with the subscript $c$ indicating conditioned on
collisions occurring at times $t_1,t_2,\ldots $, will then be
\begin{widetext}
	\begin{align}
	\rho_c(t)=&e^{\mathcal{L}_St}\theta(t_1-t)+e^{\mathcal{L}_S(t-t_1)}
	(1+\mathcal{F}_a)
	e^{\mathcal{L}_St_1}\rho(0)\theta(t_2-t)\theta(t-t_1)\notag\\
	&+e^{\mathcal{L}_S(t-t_2)}(1+\mathcal{F}_a)
	e^{\mathcal{L}_S(t_2-t_1)}(1+\mathcal{F}_a)
	e^{\mathcal{L}_St_1}\rho(0)\theta(t_3-t)\theta(t-t_2)+\ldots 
\end{align}
\end{widetext}
where $\theta(t)$ is the unit step function. Taking the derivative
with respect to time and using $\theta'(t)=\delta(t)$ we find that
$\rho(t)$ is given by the following differential equation
\begin{equation}
	\dot{\rho}_c(t)=\mathcal{L}_S\rho_c(t)+I(t)\mathcal{F}_a
\rho_c(t-\epsilon)\label{conditioned}
\end{equation}
where appears the quantity
\begin{equation}
	I(t)=\sum_{n}^{}\delta(t-t_n)
\end{equation}
sometimes referred to as generalised shot noise \cite{Rice77}.

The arrival times $t_1,t_2,\ldots $ will in general be random, so
that $I(t)$ will itself be a stochastic process. The aim then is to
derive the master equation for the density operator for the system
averaged over all realizations of $I(t)$. Thus we are seeking
$\rho(t)=\langle \langle \rho_c(t)\rangle\rangle$, for which we need
to specify the statistical properties of $I(t)$.

\subsection{Renewal process master equation\label{ReProMaEq}}

A particularly useful approach to arriving at a model for the
stochastic properties of $I(t)$ is to treat the arrival times as a
renewal process \cite{Bhat02} in which is specified a waiting time
distribution $w(\tau)$, such that $w(\tau)d\tau$ is the probability
that the next collision will occur in the time interval
$(\tau,\tau+d\tau)$ after the previous. The simplest case of a
renewal process is a Poissonian beam with mean arrival rate $R$, for
which the waiting time distribution is exponential:
\begin{equation}
	w(\tau)=R^{-1}e^{-R\tau}.
\end{equation}
In this case, averaging over the arrival times can be readily shown
to yield the Markovian master equation
\begin{equation}
	\dot{\rho}=\mathcal{L}_S\rho+R\mathcal{F}_a \rho=\mathcal{L}\rho
	\label{PoissonianResult}
\end{equation}
where $\mathcal{L}=\mathcal{L}_S+R\mathcal{F}_a$ is a Lindblad
operator.

For other choices of $w(\tau)$, the analysis relies on results of the
theory of renewal processes, and is somewhat more involved. It has
been applied in the case of the micromaser in
\cite{Cresser92b,Cresser96}, and recently in the general analysis of
the class of collisional models formulated in terms of renewal
processes in
\cite{Budini04,Budini09,Vacchini13,Vacchini14,Vacchini16}. Amongst
other results, the master equation turns out to be non-Markovian.
This has been demonstrated in the case of the micromaser in
\cite{Cresser92b,Cresser96} and again recently, though leading to
slightly different results, in \cite{Budini09,Vacchini16} for reasons
explained below. Here, a variation on these derivations alternative
to, and  more straightforward than, the derivation presented in
\cite{Cresser96}, that makes explicit use of the stationary
statistics of the shot noise $I(t)$ is presented.

The solution to the conditional master equation Eq.\ (\ref{conditioned}) can be  written as a Dyson series and the average taken over all realisations of $I(t)$ to yield the corresponding expansion for the system density operator $\rho(t)=\langle \langle
\rho_c(t)\rangle\rangle$,
\begin{widetext}
	\begin{align}
	\rho(t)=&e^{\mathcal{L}_St}\rho(0)
	+\sum_{n=1}^{\infty}\int_{0}^{t}d\tau_n
	\int_{0}^{\tau_n-\epsilon}d\tau_{n-1}\ldots
	\int_{0}^{\tau_2-\epsilon}d\tau_1
	\langle \langle I(\tau_n)I(\tau_{n-1})
	\ldots I(\tau_2)I(\tau_1)\rangle\rangle\notag\\
	&\hspace*{2cm}\times e^{\mathcal{L}_S(t-\tau_n)}\mathcal{F}_a
	e^{\mathcal{L}_S(\tau_n-\tau_{n-1})}\mathcal{F}_a \ldots
	e^{\mathcal{L}_S(\tau_2-\tau_1)}\mathcal{F}_a
	e^{\mathcal{L}_S\tau_1}\rho(0)\label{rhoExpansion}
\end{align}
\end{widetext}
where the initial time $t=0$ will, in general, lie between successive
arrivals. In this expression there appears the multitime shot noise
correlation function $\langle \langle I(\tau_n)\ldots
I(\tau_1)\rangle\rangle$.  If the stream of ancilla is assumed to
have been initiated in the infinite past, then $I(t)$ will be a
stationary stochastic process and as such will be a function of time
differences only. Further, the ranges of integration always involve
an infinitesimal offset, so that the correlation function is only
required for times satisfying the strict inequality
$t_n>t_{n-1}>\ldots >t_2>t_1$. In that case, a singular value of the
correlation function for equal time arguments will not contribute,
and it can be shown \cite{Rice77,Benkert90} that, on averaging over
the collision times, the correlation function is given by
\begin{multline}
	\langle\langle I(\tau_n)I(\tau_{n-1})\ldots
	I(\tau_2)I(\tau_1)\rangle\rangle\\
	=R^{n}g(\tau_n-\tau_{n-1})g(\tau_{n-1}-\tau_{n-2})\ldots 
	g(\tau_2-\tau_1)\\ t_n>t_{n-1}>\ldots >t_1
	\label{ShotNoiseCorrelation}
\end{multline}
where $g(\tau)=R^{-2}\langle \langle I(t)I(t+\tau)\rangle\rangle$ is
a normalized `intensity' correlation function for the incident
ancilla, a function of time differences only, as expected for a
stationary process. In this expression $R=\langle \langle
I(t)\rangle\rangle$ is the mean collision rate and is given by
\begin{equation}
	R^{-1}=\int_{0}^{\infty}\tau w(\tau)d\tau
\end{equation}
while $g(t)$, the normalised intensity correlation function, also
known as the renewal density \cite{Rice77,Bhat02}, or sprinkling
distribution, satisfies the integral equation
\begin{equation}
	Rg(t)=w(t)+R\int_{0}^{t}w(\tau)g(t-\tau)ds.
	\label{gDefined}
\end{equation}
Eq.\ (\ref{ShotNoiseCorrelation}) can be substituted into Eq.\
(\ref{rhoExpansion}) from which, as shown in \ref{MEDerivation}, can be derived the master equation for $\rho(t)$
\cite{Cresser92b,Cresser96}
\begin{equation}
	\dot{\rho}=\mathcal{L}_S\rho(t)+R\mathcal{F}_a
\int_{0}^{t}e^{\mathcal{L}_S(t-\tau)}\mathcal{K}(t-\tau)\rho(\tau)d\tau
	\label{nonMarkovME}
\end{equation}
which, by virtue of the appearance of a memory kernel $\mathcal{K}(t)$, describes a generally non-Markovian evolution. The Laplace transform of the memory kernel $\mathcal{K}(t)$ is
\begin{equation}
	\tilde{\mathcal{K}}(s)=\left(1-(\tilde{g}(s)-s^{-1})R\mathcal{F}_a
\right)^{-1}.
	\label{LaplaceMemoryKernel}
\end{equation}
For Poissonian statistics, $g(t)=1$, the memory Kernel becomes a
delta function, $\mathcal{K}(t)=\delta(t)$ and the master equation
reduces to the standard result Eq.\ (\ref{PoissonianResult}).

The master equation derived above can be contrasted with that found
in, e.g., \cite{Vacchini16}, (and in particular in the supplement
\cite{VacchiniSupp16}) where an analysis is made on the basis of slightly
different assumptions concerning the implementation of the renewal process
description of the interaction times. These concern the choice of the
exclusive probability densities for jumps corresponding to the action
of $(1+\mathcal{F}_a)$ at times $t_1, t_2\ldots$, the difference in
outcome lying in the early $t$ dependence of $\rho(t)$ on the initial
state $\rho(0)$. A comparison can be made by taking the Laplace transform of Eq.\
(\ref{nonMarkovME}) and rearranging terms, so that this equation can be cast
in the form
\begin{align}
	\dot{\rho}=&\mathcal{L}_S\rho+R\mathcal{F}\int_{0}^{t}
	e^{\mathcal{L}_S(t-\tau)}k(t-\tau)\rho(\tau)d\tau\notag\\
	&-R\mathcal{F}_a
	e^{\mathcal{L}_St}(g(t)-1)\rho(0)
	\label{NearlyVacchini}
\end{align}
where $\tilde{k}(s)=s\tilde{g}(s)$. For sufficiently large $t$,
$g(t)\to1$, so the term on the right hand side depending on the
initial state $\rho(0)$ will become negligible and we are left with
the result derived in \cite{VacchiniSupp16} (which was incorrectly
stated as being the result found in \cite{Cresser96}),
\begin{equation}
	\dot{\rho}=\mathcal{L}_S\rho+R\mathcal{F}
	\int_{0}^{t}e^{\mathcal{L}_S(t-\tau)}k(t-\tau)\rho(\tau)d\tau.
	\label{VacchiniMEq}
\end{equation}
The reason for this is best understood by considering the quantum
trajectory unravelling of $\rho(t)$.

\subsection{Quantum trajectory unravelling\label{UnravelledNonMark}}

The appearance of a non-Markovian master equation belies the fact that
the dynamics possesses a straightforward quantum trajectory
unravelling. This is because the non-Markovianity
has its origins in an externally imposed source of noise, in contrast
to non-Markovianity that arises in systems coupled to a quantum
reservoir, where a `back-flow' of information produced by the
system-reservoir dynamics is the underlying cause of the non-Markov
behaviour. 

The required unravelling can be obtained by first substituting Eq.\ (\ref{ShotNoiseCorrelation}) into Eq.\ (\ref{rhoExpansion}), doing so yielding a series expansion which, in spite of appearances, is in fact not a possible quantum jump unravelling as $\mathcal{F}_a $ is not a jump operator (the trace of its output is zero). But after some reorganisation of terms, as shown in \ref{Unravelling}, the required expansion expressed in terms of the jump operator $1+\mathcal{F}_a$ can be arrived at
\begin{widetext}
	\begin{align}
	\rho(t)=&e^{\mathcal{L}_St}
	p_0(t)\rho(0)+\sum_{n=1}^{\infty}\int_{0}^{t}dt_n\int_{0}^{t_n}dt_{n-1}\ldots
	\int_{0}^{t_2}dt_1p_f(t-t_n)w(t_n-t_{n-1})\ldots
	w(t_2-t_1)p_1(t_1)\nonumber\\ &\times
	e^{\mathcal{L}_S(t-t_n)}(1+\mathcal{F}_a)
	e^{\mathcal{L}_S(t_n-t_{n-1})}(1+\mathcal{F}_a )\ldots
	(1+\mathcal{F}_a )e^{\mathcal{L}_St_1}\rho(0)
	\label{DysonLike}
\end{align}
\end{widetext}
where
\begin{equation}
	p_f(t-t_n)=1-\int_{0}^{t-t_n} w(\tau)d\tau
\end{equation}
is the probability that there is no collision in the time interval
$(t_n,t)$ after the final collision at time $t_n$,
\begin{equation}
	p_1(t_1)dt=p_f(t_1)Rdt_1
	\label{pResid}
\end{equation}
known as the residual time probability distribution, is the
probability that no collision occurs in the time interval $(0,t_1)$,
and a collision, the first after $t=0$, occurs in the interval
$(t_1,t_1+dt_1)$, and
\begin{equation}
	p_0(t)=1-\int_{0}^{t}p_1(\tau)d\tau
\end{equation}
is the probability that there are no collisions in the time interval
$(0,t)$. That the probabilities $p_0(t)$ and $p_1(t)$ take the form that they
do is a consequence of the shot noise $I(t)$ being stationary.
Starting the evolution at an arbitrary initial time $t=0$ entails
introducing the residual time probability for the arrival of the
first ancilla after $t=0$, given as above by $p_1(t)$ \cite{Bhat02}. 

Any unravelling into an ensemble of quantum trajectories can then be
carried out by simulating the ancilla interaction times according to
the set of probabilities $p_0(t), p_1(t), w(t)$ and $p_f(t)$. 

The above result was derived starting from the form Eq.\ (\ref{ShotNoiseCorrelation}) for the shot noise correlation. An alternate procedure is to construct the exclusive probability for a sequence of collisions, here given by the product in Eq.\ (\ref{DysonLike}), $p_f(t-t_n)w(t_n-t_{n-1})\ldots w(t_2-t_1)p_1(t_1)$. The master equation of \cite{VacchiniSupp16}, Eq.\
(\ref{VacchiniMEq}), can then be shown to arise if we set $p_1(t)=w(t)$, i.e., the waiting time distribution $p_1(t)$ for the arrival of the first ancilla after the initial time $t=0$ is replaced by the intercollision waiting time distribution $w(t)$. Thus this master equation corresponds to a different modeling of the statistics of the arrival times -- one in which the the initial system state is set at a time immediately after a collision -- but which only has an impact for short times, after which Eq.\ (\ref{NearlyVacchini}) reduces to Eq.\ (\ref{VacchiniMEq}).

\subsection{Relaxation to steady state}

The steady state solution to the general master equation Eq.\ (\ref{nonMarkovME}), achieved when $\dot{\rho}=0$, plays an essential role in determining the
time-reversal quantum trajectory properties of the system, required in the discussion later in Section \ref{TRQT} of fluctuation relations.
This steady state is most easily determined by working with the Laplace transform of Eq.\ (\ref{nonMarkovME}), and using $\rho(\infty)=\rho_{ss}=\lim_{s\to0}s\tilde{\rho}(s)$ is given by the solution of
\begin{equation}
	\mathcal{L}_S\rho_{ss}+R\mathcal{F}_a
	\tilde{g}(-\mathcal{L}_S)(-\mathcal{L}_S)\rho_{ss}=0.
\end{equation}
For Poissonian statistics $(g=1)$ the above reduces to 
\begin{equation}
	\mathcal{L}_S\rho_{ss}+R\mathcal{F}_a\rho_{ss}=0.
	\label{PoissonSteadyState}
\end{equation}
As an example of the steady state solution for non-Poissonian
statistics, we can consider the case of
\begin{equation}
	g(t)=Ae^{-\Gamma t}+1\label{SuperBunchedg}
\end{equation}
which is the renewal density (intensity correlation) for
super-bunched (for $A>1$) ancilla interactions. In this case, in the
limit of $\Gamma\to0$, the steady state can be readily shown to be
\begin{equation}
	\rho_{ss}=(1+A)^{-1}(\rho_A+A\rho_{eq})
	\label{SuperBunchedrhoss}
\end{equation}
where $\rho_{eq}$ is the equilibrium state for the cavity in the
absence of any ancilla, while $\rho$ is the steady state solution to
\begin{equation}
	\mathcal{L}_S\rho_A+AR\mathcal{F}_a\rho_A=0
\end{equation}
i.e., the steady state for the system with a Poissonian interaction
rate $AR$. This result is easy to understand: excitations by the
ancilla will occur in Poissonian bursts at an enhanced rate $AR$,
separated by quiescent intervals in which the cavity will relax to
its equilibrium state.

\section{The micromaser\label{micromaser}}

The micromaser is an early example of a collisional or repeated
interaction model that has recently been analyzed from the
perspective of its thermodynamic properties \cite{Strasberg17}. For
the micromaser, the system is a single-mode cavity field of frequency
$\omega_S$ which interacts with a succession of qubits (highly
excited Rydberg atoms) with a transition frequency $\omega_a$ near or
on resonance with the cavity field frequency. These atoms are
typically, but not necessarily, prepared in a fully inverted state.
Between qubit interactions, the cavity is weakly damped by coupling
to its thermal environment of temperature $T_{c}$. The system
Hamiltonian is then $H_S=\omega_S a^\dagger a$ while the effect of the coupling to the external environment is described by the usual Lindblad form
\begin{align}
	\mathcal{L}_S\rho=&-i\omega_S\left[a^\dagger a,\rho\right]\notag\\
	&+(\bar{n}+1)\gamma\left(a\rho a^\dagger-\tfrac{1}{2}
	\left\{a^\dagger
	a,\rho\right\}\right)\notag\\
	&+\bar{n}\gamma\left(a^\dagger\rho
	a-\tfrac{1}{2}\left\{a a^\dagger ,\rho\right\}\right)
\end{align}
where $\bar{n}=\big(e^{\hbar\omega_a/kT_\text{c}}-1\big)^{-1}$.

The ancilla are qubits with Hamiltonian $H_a=\tfrac{1}{2}\omega_a\sigma_z$
that interact with the cavity field for a brief period $\tau_\text{int}$, this interaction being described by the Jaynes-Cummings interaction $V=\Omega\left(\sigma_-a^\dagger+\sigma_+a\right)$.
The prepared state of the qubits will be assumed to be diagonal in
their energy basis $\rho_a=p|e\rangle\langle e|+(1-p)|g\rangle\langle g|$.
In the original model for the micromaser, the qubits were assumed to
be fully inverted, $p=1$, but in general they will be taken to have
exited from a thermal reservoir of temperature $T_a$, so that
$p=\big(e^{\hbar\omega_a/kT_a}+1\big)^{-1}$.

Assuming exact resonance between the qubit and the cavity field,
$\omega_a=\omega_S=\omega$, the interaction of the $n$th qubit with
the cavity is described by Eq.\ (\ref{AncillaJump})
where, in the impulsive limit $\tau_\text{int}\to0$ and
$\Omega\to\infty$ with $\theta=\Omega\tau_\text{int}$ held fixed, we
find that the Lindbald operators, from Eq.\
(\ref{LindbladSystemAncilla}), $L_{mn}=\sqrt{p_m}\langle
m|\exp(-iV\tau)|n\rangle$, $m,n=e,g$ associated with the interaction
of the cavity field with an incident atom are 
\begin{equation}
	\begin{split}
		&L_{ee}=\sqrt{p}\cos(\theta\sqrt{N+1})\qquad
		L_{ge}=\sqrt{p}\,\frac{\sin(\theta\sqrt{N})}{\sqrt{N}}a^\dagger\\
		&L_{gg}=\sqrt{1-p}\,\cos(\theta\sqrt{N})\quad
		L_{eg}=\sqrt{1-p}\,\frac{\sin(\theta\sqrt{N+1})}{\sqrt{N+1}}a.
	\end{split}
\end{equation}
In terms of these operators $(1+\mathcal{F}_a)\rho$ is
\begin{align}
	\left(1+\mathcal{F}_a\right){\rho}
	=&L_{ee}\rho L_{ee}^\dagger+L_{ge}\rho L_{ge}^\dagger\notag\\
	&+L_{gg}\rho L_{gg}^\dagger+L_{eg}\rho L_{eg}^\dagger
\end{align}
The operators $L_{mn}\sqrt{dt}$ are the Kraus operators corresponding
to measurements made on the qubits exiting the cavity. Thus. for
instance, $L_{ee}\rho L_{ee}^\dagger$ is a mapping of the state of
the cavity conditioned on an incident qubit in its excited state
being measured to be in its excited state on exiting the cavity,
while $L_{eg}\rho L_{eg}^\dagger$ is a mapping of the cavity state
conditioned on an incident qubit in its ground state being measured
to be in its excited state, with a photon thereby having been
absorbed from the cavity field. These `quantum jumps' will occur with
a probability $\text{Tr}_a[L_{mn}\rho L_{mn}^\dagger]dt$ in the time
interval $(t,t+dt)$.

Any realisation of the measurements implied by these operators will
require either a measurement of the qubit state (whether excited or
ground) prior to interacting with the cavity field, followed by a
measurement after the interaction ceases, or by assuming that there
are in fact two distinguishable beams, one of excited atoms, the
other of ground state atoms, the first arriving at a rate $pR$, the
second at a rate $(1-p)R$, which each beam subject to separate
measurements after interaction has ceased.

\subsection{Micromaser master equation}

It is typically the case that an exponential waiting time
distribution is assumed for the incident atoms, in which case the
master equation reduces to the form Eq.\ (\ref{PoissonianResult}),
and is given by
	\begin{equation}
	\begin{split}
		\dot{\rho}=&\mathcal{L}_S\rho+L_{ee}\rho L_{ee}^\dagger-\tfrac{1}{2}\left\{L_{ee}^\dagger L_{ee},\rho\right\}\\
		&+L_{ge}\rho L_{ge}^\dagger-\tfrac{1}{2}\left\{L_{ge}^\dagger L_{ge},\rho\right\}\\
		&+L_{gg}\rho L_{gg}^\dagger-\tfrac{1}{2}\left\{L_{gg}^\dagger L_{gg},\rho\right\}\\
		&+L_{eg}\rho L_{eg}^\dagger-\tfrac{1}{2}\left\{L_{eg}^\dagger L_{eg},\rho\right\}
	\end{split}
\end{equation}
a result first obtained, for $p=1$, i.e., where the incident atoms
are all fully excited, by \cite{Filipowicz86, Lugiato87}.

\subsection{Weak coupling limit}

In the above master equation, the coupling to the atomic beam
reservoir is not assumed weak, in contrast to the usual weak coupling
assumption made in deriving Markov master equations. The weak
coupling limit of $\theta\ll1$ is nevertheless revealing. Provided
the mean photon numbers in the cavity are not too high, this becomes,
with $\bar{n}_a=p\theta^2=(e^{\hbar\omega/kT_a}-1)^{-1}$
\begin{align}
	\dot{\rho}=&-i \omega \left[a^\dagger a,\rho\right]
	+(\bar{n}+1)\gamma\left(a\rho a^\dagger
	-\tfrac{1}{2}\left\{a^\dagger a,\rho\right\}\right)\notag\\
	&+\bar{n}\gamma\left(a^\dagger\rho a
	-\tfrac{1}{2}\left\{aa^\dagger,\rho\right\}\right)\notag\\
	&+\bar{n}_aR\left(a^\dagger\rho a
	-\tfrac{1}{2}\left\{aa^\dagger,\rho\right\}\right)\notag\\
	&+(\bar{n}_a+1)R\left(a\rho a^\dagger
	-\tfrac{1}{2}\left\{a^\dagger a,\rho\right\}\right)
	\label{ThermalLimit}
\end{align}
indicating the beam acts as a thermal reservoir at the temperature
$T_a$ of the beam atoms, and has been commonly used in this fashion,
see e.g., \cite{Sargent74,Kist99,Horowitz12}.

This result holds true even in the case of non-exponential waiting
times. In that case, since $\mathcal{F}_a\sim\theta^2$, the Laplace
transform of the memory kernel $\tilde{K}(s)$, Eq.\
(\ref{LaplaceMemoryKernel}) can be replaced by unity, so that
$K(t)\sim\delta(t)$ and the non-Markovian master equation reduces to
Eq.\ (\ref{PoissonianResult}) from which Eq.\ (\ref{ThermalLimit})
follows again.

\subsection{Steady State}

The steady state of the cavity field $\rho_{ss}=\rho(\infty)$ will be
given by Eq.\ (\ref{PoissonSteadyState}). It is diagonal in the
number basis,
\begin{equation}
	\rho_{ss}=\sum_{n=0}^{\infty}P_{ss}(n)|n\rangle\langle n|
\end{equation}
where the probability of finding $n$ photons in the cavity at steady
state, $P_{ss}(n)$, is given by
\begin{equation}
P_{ss}(n)=P_{ss}(0)\prod_{m=1}^n\frac{pR\sin^2\big(\!\sqrt{m}\theta\big)/m+\gamma\bar{n}}
{(1-p)R\sin^2\big(\!\sqrt{m}\theta\big)/m+\gamma(\bar{n}+1)}
\label{SteadyStatePhotonNumber}
\end{equation}
and with $P_{ss}(0)$ determined by the requirement that
$$\sum_{n=0}^{\infty}P_{ss}(n)=1.$$

\subsection{Quantum trajectory analysis\label{QuTrajMicro}}

The master equation for the micromaser is of Lindblad form, and so is
amenable to standard quantum trajectory analysis
\cite{Cresser96,Casagrande98,Chough13}. 

Introducing the operators $C_{-1}$ and $D_{-1}$
\begin{equation}
	\begin{split}
		C_{-1}=&\sqrt{(\bar{n}+1)\gamma}a\\
		D_{-1}=&\sqrt{(1-p)R}
		\frac{\sin\big(\theta\sqrt{N+1}\big)}{\sqrt{N+1}}a
	\end{split}
\end{equation}
which represent a loss of a photon from the cavity to the reservoir,
or absorbed by an atom respectively, and
\begin{equation}
	\begin{split}
		C_1=&\sqrt{pR}\frac{\sin\big(\theta\sqrt{N}\big)}
		{\sqrt{N}}a^\dagger\\
		D_1=&\sqrt{\bar{n}\gamma}a^\dagger
\end{split}
\end{equation}
which represent a gain of a photon from the cavity reservoir, or from an atom respectively, and finally, for convenience, a further pair of operators associated with the atom passing through the cavity without giving up or absorbing a photon are defined by
\begin{equation}
	\begin{split}
		C_0=&\sqrt{pR}\cos\big(\theta\sqrt{N+1}\big)\\
		D_0=&\sqrt{(1-p)R}\cos\big(\theta\sqrt{N}\big),
	\end{split}
\end{equation}
we can write the master equation as
\begin{align}
	\dot{\rho}=&-i \omega \left[a^\dagger a,\rho\right]
	+\sum_{i=-1}^{1}\left[C_i \rho C_i^\dagger 
	-\tfrac{1}{2}\left\{C_i^\dagger  C_i ,\rho\right\}\right.\notag\\
	&\left.+D_i \rho D_i^\dagger 
	-\tfrac{1}{2}\left\{D_i^\dagger  D_i ,\rho\right\}\right].
	\label{FullMicromaserME}
\end{align}
The dynamics of the micromaser can then be unravelled in terms of the jump operators $\mathcal{J}_i$ and $\mathcal{K}_i$ defined by
\begin{equation}
	\mathcal{J}_i\rho=C_i\rho C_i^\dagger\quad\text{and}\quad
\mathcal{K}_i\rho=D_i\rho D_i^\dagger
\end{equation}
with the jumps for $i=\pm1$ representing the gain or loss of a single
photon from the cavity field, and $i=0$ representing no change in the
cavity field photon number. Between jumps the system evolution is
determined by the non-Hermitean Hamiltonian
\begin{equation}
	H_c= \omega a^\dagger a-\tfrac{1}{2}i\sum_{i}^{}\left[C_i^\dagger
C_i+D_i^\dagger D_i\right]\label{Hc}
\end{equation}
with the between-jumps evolution given by the superoperator
$\mathcal{L}_c$:
\begin{equation}
	\mathcal{L}_c\rho=-i\left[H_c\rho-\rho H_c^\dagger\right].
\end{equation}
A Dyson series decomposition of the full dynamics then reads
\begin{widetext}
	\begin{equation}
	\begin{split}
		\rho(t)
		=\rho_c^{(0)}(t)P^{(0)}(t)
		+&\sum_{n=1}^{\infty}\sum_{\mathcal{L}_{i_1}}
		\sum_{\mathcal{L}_{i_2}}\ldots \sum_{\mathcal{L}_{i_n}}
		\int_{0}^{t}dt_n\int_{0}^{t_n}dt_{n-1}\ldots \int_{0}^{t_2}dt_1\\
		&\times P^{(n)}(t;\mathcal{L}_{i_1},t_1,\ldots,
		\mathcal{L}_{i_n},t_n)\rho_c^{(n)}(t;
		\mathcal{L}_{i_1},t_1,\ldots, \mathcal{L}_{i_n},t_n)
	\end{split}
\end{equation}
where the state of the cavity field at time $t$ conditioned on the
sequence of jumps
$\mathcal{L}_i\in\left\{\mathcal{J}_i,\mathcal{K}_i\right\}$
occurring at times $t_1,t_2,\ldots ,t_n$ is
	\begin{equation}
	\rho_c^{(n)}(t;\mathcal{L}_{i_1},t_1,\ldots, \mathcal{L}_{i_n},t_n)
=\frac{e^{\mathcal{L}_c(t-t_n)}\mathcal{L}_{i_n}e^{\mathcal{L}_c(t_{n}-t_{n-1})}\mathcal{L}_{i_{n-1}}\ldots
\mathcal{L}_{i_1}e^{\mathcal{L}_ct_1}\rho(0)}{\text{Tr}[e^{\mathcal{L}_c(t-t_n)}\mathcal{L}_{i_n}e^{\mathcal{L}_c(t_{n}-t_{n-1})}\mathcal{L}_{i_{n-1}}\ldots
\mathcal{L}_{i_1}e^{\mathcal{L}_ct_1}\rho(0)]}.
\end{equation}
A quantum trajectory $\gamma$ can then be defined as a sequence of
states
$$\gamma\equiv\{\rho_c^{(0)}(t),\rho_c^{(1)}(t;\mathcal{L}_{i_1},t_1),\rho^{(2)}_c(t;\mathcal{L}_{i_2},t_2,\mathcal{L}_{i_1},t_1),\ldots
\}$$
\end{widetext} conditioned on the sequence of measurements implied by the jump
operators $\mathcal{L}_i$. Such a trajectory occurs with a
probability $P^{(n)}[\gamma](dt)^n$ and is given by the trace of the
final state of the sequence, i.e.
\begin{multline}
P^{(n)}[\gamma]=P^{(n)}(t;\mathcal{L}_{i_1},t_1,\ldots,
		\mathcal{L}_{i_n},t_n)
		\\=\text{Tr}[e^{\mathcal{L}_c(t-t_n)}\mathcal{L}_{i_n}e^{\mathcal{L}_c(t_{n}-t_{n-1})}\mathcal{L}_{i_{n-1}}\ldots
\mathcal{L}_{i_1}e^{\mathcal{L}_ct_1}\rho(0)].\label{QTProb}
\end{multline}
If the initial state $\rho(0)$ is a pure state,
$\rho(0)=|\psi(0)\rangle\langle \psi(0)|$, and since the jump
operators map pure states into pure states, the quantum trajectory
can be written as a sequence of pure states,
$|\psi_c^{(n)}(t)\rangle$, with the probability of a trajectory
occurring then given by the norm $\langle
\psi_c^{(n)}(t)|\psi_c^{(n)}(t)\rangle$.

\section{Time reversed quantum trajectories\label{TRQT}}

Time reversed quantum trajectories have gained significant attention
in recent times in the context of understanding fluctuation theorems
of statistical mechanics in a quantum setting. The original classical
fluctuation theorems \cite{Crooks99,Evans02} relate the probabilities
to observe particular classical microscopic trajectories related by
time reversal and typically take the form
\begin{equation}
	\frac{p_F(x)}{p_R(-x)}=\exp\left[a(x-b)\right]
	\label{FluctuationTheorem}
\end{equation}
where $x$ can be, for instance, entropy produced or energy (heat)
transported, $p_F(x)$ is the probability of amount $x$ being
transported in the `forward' direction, and $p_R(-x)$ is the
probability transported in the `backward' direction. These theorems
have been extended into the quantum regime,
\cite{Horowitz12,Horowitz13,Hekking13,Elouard17} with the role of the
microscopic classical trajectories played by quantum trajectories.

The essential idea in constructing a time-reversed quantum trajectory
lies in associating with any trajectory in the forward time
direction, a conjugate trajectory in the time reversed direction. It
serves to refine the notion of a quantum trajectory at this point,
which is to assume that the initial state of the the forward process
is an eigenstate of some observable, which in the case of the
micromaser will invariably be an eigenstate of the photon number
operator $N$, $|n\rangle$ say. A series of $k$ quantum jumps
$\mathcal{L}_i$ interleaved with no-jump non-Hermitean evolution
generated by $\mathcal{L}_c$ is then projected onto the final state
$|m\rangle$ at time $t$. If we adopt the notation
$\gamma_{nm}\equiv\{n,0;\mathcal{L}_1,t_1;\ldots
\mathcal{L}_{k-1},t_{k-1};\mathcal{L}_k,t_k;m,t\}$ (with time
increasing from left to right, opposite to how it occurs in the
expression Eq.\ (\ref{QTProb})) to represent such a quantum
trajectory then the time reversed quantum trajectory $\tilde{\gamma}$
is then taken to start at the time reversed state
$|\tilde{m}\rangle=\Theta|m\rangle$ where $\Theta$ is the time
reversal operator, and end at time $t$ in the time reversed state
$|\tilde{n}\rangle$:
\begin{equation}
\tilde{\gamma}_{\tilde{m}\tilde{n}}=\{\tilde{m},0;\tilde{\mathcal{L}}_k;t-t_k;\ldots;\tilde{\mathcal{L}}_{2},t-t_2;\tilde{\mathcal{L}}_1,t-t_1;\tilde{n},t\}
\end{equation}
where the $\tilde{x}$ indicate that the time reversed counterparts of
$x$ are to be inserted. These conjugate trajectories will then occur
with the conditional probability $P[\gamma_{nm}]$, i.e., conditioned
on the initial state being $|m\rangle$ for the forward trajectory,
and $\tilde{P}[\tilde{\gamma}_{\tilde{m}\tilde{n}}]$ the conditional
probability for the backward trajectory.

For the micromaser, the cavity field Hamiltonian $H_S=\omega
a^\dagger a$ is time reversal invariant, so we can set
$\Theta|m\rangle=|\tilde{m}\rangle=|m\rangle$, and since from Eq.\
(\ref{Hc}), $\Theta iH_c\Theta^\dagger=-iH_c^\dagger$, we have
\begin{equation}
	\tilde{\mathcal{L}}_c\rho
	=i\left[H_c\rho-\rho H_c^\dagger\right]
\end{equation}
which leaves the task of determining the form for the time reversed
jump operators $\tilde{\mathcal{L}}_i$.

The notion of a time-reversed quantum trajectory is not unique, this
arising through the different approaches to defining the time
reversed jump operators \cite{Horowitz13}. The various possibilities
that have been proposed can be shown to be closely related to one
another \cite{Manzano15} and in particular to one introduced by
Crooks\cite{Crooks08}, which is based, for systems that have reached
steady state, on imposing a time symmetric condition on the
probabilities for a forward and its time-reversed trajectory. If a
trajectory $\gamma$ specified by a sequence of jumps in the forward
direction, starting in the steady state occurs with probability
$P[\gamma]$, then a time reversed \emph{dual} trajectory
$\tilde{\gamma}$ is then that trajectory involving a sequence of
jumps in the reversed direction for which the steady state
probability $\tilde{P}[\tilde{\gamma}]$ of observing $\tilde{\gamma}$
is the same as observing the trajectory $\gamma$ in the original
process: $\tilde{P}[\tilde{\gamma}]=P[\gamma]$ \cite{Crooks08}.
Crooks shows that this leads to the following prescription for
constructing the time reversed (dual) jump operators
\begin{equation}
\tilde{\mathcal{L}}_i=\Theta\rho_{ss}^{1/2}\mathcal{L}_i^\dagger\rho_{ss}^{-1/2}\Theta^{-1}
	\label{CrooksTR}
\end{equation}
where $\rho_{ss}$ is the steady state density operator, as given by
Eq.\ (\ref{SteadyStatePhotonNumber}) for the micromaser cavity field.
The above expression is that which was originally derived by Crooks,
though without the pre and post factors $\Theta\ldots \Theta^{-1}$.

This idea is developed further below for application to the
micromaser, but first it is shown how the condition arises in a
natural fashion under certain circumstances relating cavity field and
atomic beam correlation functions.

\section{Dual quantum trajectories for the
micromaser\label{DualQuaTraj}}
 
In the case of the micromaser the notion of a dual trajectory has a
direct physical interpretation in that instance in which the atoms in
the incoming atomic beam are fully excited, $p=1$, and the cavity
field reservoir is at zero temperature. In that case, the master
equation Eq.\ (\ref{FullMicromaserME}) reduces to
\begin{equation}
	\dot{\rho}=-i \omega \left[a^\dagger
a,\rho\right]+\sum_{i=-1}^{1}\left[C_i\rho
C_i^\dagger-\tfrac{1}{2}\left\{C_i^\dagger C_i,\rho\right\}\right]
\end{equation}
with the steady state given by
$\rho_{ss}=\sum_{n}P_{ss}(n)|n\rangle\langle n|$, with $P_{ss}(n)$
from Eq.\ (\ref{SteadyStatePhotonNumber}) with $\bar{n}=0$ and $p=1$:
\begin{equation}
	P_{ss}(n)
	=P_{ss}(0)\left(\frac{R}{\gamma}\right)^n
	\prod_{m=1}^n\frac{\sin^2\big(\!\sqrt{m}\theta\big)}{m}
	\label{SpecialEquil}
\end{equation}
with
\begin{equation}
	C_{-1}=\sqrt{\gamma}a,\qquad
C_1=\sqrt{R}\frac{\sin\big(\theta\sqrt{N}\big)}{\sqrt{N}}a^\dagger.
\end{equation}
The steady state is time reversal invariant,
$\Theta\rho_{ss}\Theta^{-1}=\rho_{ss}$, so that $P_{ss}(n)=\langle
n|\rho_{ss}|n\rangle=\langle
\tilde{n}|\rho_{ss}|\tilde{n}\rangle=\tilde{P}_{ss}(n)$.

For the micromaser, the relevant probability will be the probability
of measuring $m$ photons in the cavity at time $t=0$ (when the cavity
field is already at steady state) and $n$ photons in the cavity at a
time $t$ later for a particular quantum trajectory $\gamma$. This
probability, $P_{nm}[\gamma]$ is given by
\begin{widetext}
	\begin{equation}
	\begin{split}
		P_{nm}[\gamma]
		=&\left\lvert \langle n|
		e^{-iH_c(t-t_k)}C_{i_k}e^{-iH_c(t_k-t_{n-1})}C_{i_{n-1}}\ldots
C_{i_1}
		e^{-iH_ct_1}|m\rangle\right\rvert^2P_{ss}(m)\\
		=&\left\lvert \langle n|\rho_{ss}^{-1/2}e^{-iH_c(t-t_k)}I 
		C_{i_k}I\ldots I C_{i_1}I e^{-iH_ct_1}\rho_{ss}^{1/2}
		|m\rangle\right\rvert^2P_{ss}(n)
	\end{split}
\end{equation}
where the unit operator $I$ has been introduced between neighbouring
operator factors. We now proceed by inserting
$\rho_{ss}^{1/2}\rho_{ss}^{-1/2}=I$ and making use of
$\left[H_c,\rho_{ss}\right]=0$ to yield
\begin{equation}
	P_{nm}[\gamma]=\left\lvert \langle
n|Ie^{-iH_c(t-t_k)}I\rho_{ss}^{-1/2}C_{i_k}\rho_{ss}^{1/2}I\ldots
I\rho_{ss}^{-1/2} C_{i_1}\rho_{ss}^{1/2}I
e^{-iH_ct_1}I|m\rangle\right\rvert^2P_{ss}(n).
\end{equation}
\end{widetext}
Substituting the decomposition of the unit operator in terms of the
time reversal operator $\Theta$, $\Theta^{-1}\Theta=I$, using $\Theta
iH_c^\dagger\Theta^{-1}=-iH_c$ and making the substitution from the
Crooks definition, Eq.\ (\ref{CrooksTR}),
$\tilde{C}_i^\dagger=\Theta\rho_{ss}^{-1/2}
C_{i}\rho_{ss}^{1/2}\Theta^{-1}$ then yields
\begin{equation}
	P_{nm}[\gamma]=\left\lvert \langle
n|\Theta^{-1}e^{iH_c^\dagger(t-t_k)}\tilde{C}_{i_k}^\dagger\ldots
\tilde{C}_{i_1}^\dagger e^{iH_c^\dagger
t_1}\Theta|m\rangle\right\rvert^2P_{ss}(n)
\end{equation}
In terms of the time reversed states
$|\tilde{m}\rangle=\Theta|m\rangle$, and using $\langle
\beta|\Theta^{-1} A\Theta|\alpha\rangle=\langle
\tilde{\alpha}|A^\dagger|\tilde{\beta}\rangle$ this then is
\begin{equation}
	P_{nm}[\gamma]=\left\lvert \langle \tilde{m}|
	e^{-iH_ct_1}\tilde{C}_{i_1}\ldots \tilde{C}_{i_k} 
	e^{-iH_c (t-t_k)}|\tilde{n}\rangle\right\rvert^2
	\tilde{P}_{ss}(n).
\end{equation}
Since the cavity field Hamiltonian $H_S=\omega a^\dagger a$ is time
reversal invariant we can set for the eigenstates $|n\rangle$,
$\Theta|n\rangle=|\tilde{n}\rangle=|n\rangle$. Further by making the
substitutions $t_l\to t-t_{k-l}$ to reverse the time order, we have
the following probability for the time reversed trajectory
$\tilde{\gamma}$:
\begin{equation}
	P_{mn}[\tilde{\gamma}]=\left\lvert \langle m|
	e^{-iH_c(t-t_k)}\tilde{C}_{i_1}\ldots \tilde{C}_{i_k} 
	e^{-iH_c t_1}|n\rangle\right\rvert^2
	P_{ss}(n)
	\label{DualEquality}
\end{equation}
which is the same probability as the forward trajectory $\gamma$,
$P_{nm}[\gamma]$.

Using Eq.\ (\ref{SpecialEquil}), it readily follows that
$\tilde{C}_i=C_{-i}$: the jump operator that adds a photon to the
cavity field $C_1$ through the de-excitation of an atom, is mapped
into a jump operator that removes a photon from the field, $C_{-1}$,
through loss to the cavity reservoir, and vice versa, while $C_0$ is
left unchanged. Thus we have
\begin{multline}
	P_{mn}[\tilde{\gamma}]=\left\lvert \langle m|
	e^{-iH_c(t-t_k)}C_{-i_1}\ldots C_{-i_k} 
	e^{-iH_c t_1}|n\rangle\right\rvert^2
	P_{ss}(n)\\
	=P_{nm}[\gamma]
	\label{TRProbab}
\end{multline}
a result made use of below.

\subsection{Field and atomic beam correlation\label{FieldBeamCorr}}

The significance of this result lies in the fact that the cavity
field intensity correlation function, $g^{(2)}(t)$, given by
\begin{align}
	g^{(2)}(t)=&\frac{\langle a^\dagger(0)
a^\dagger(t)a(t)a(0)\rangle}{\langle a^\dagger
a\rangle^2}\notag\\
=&\frac{N_{ex}}{\langle a^\dagger a\rangle^2}
	\sum_{m,n=0}^{\infty}\sin^2(\theta\sqrt{n+1})m P(n,t;m,0)
\end{align}
and the correlation function for the detection of ground state atoms
emerging from the cavity, $g_1(t)$
\begin{align}
	g_1(t)=&\frac{\langle
C_1^\dagger(0)C_1^\dagger(t)C_1(t)C_1(0)\rangle}{\langle C_1^\dagger
C_1\rangle^2}\notag\\
=&\frac{N_{ex}}{\langle a^\dagger a\rangle^2}
	\sum_{m,n=0}^{\infty}\sin^2(\theta\sqrt{n+1})m P(m,t;n,0)
\end{align}
depend on the total probability $P(m,t;n,0)$ of observing $m$ photons
in the cavity at time $t$ given that $n$ were observed at time $0$ as
a generalised sum over the probabilities $P_{nm}[\gamma]$ for all the
quantum trajectories connecting the initial state $|m\rangle$ to the
final state $|n\rangle$:
\begin{widetext}
	\begin{equation}
	\begin{split}
		P(n,t;m,0)=&\sum_{\gamma}^{}P_{nm}[\gamma]\\
		=&\sum_{k=0}^{\infty}
		\sum_{i_1=-1}^{1}\sum_{i_2=-1}^{1}\ldots \sum_{i_k=-1}^{1}
		\int_{0}^{t}dt_k\int_{0}^{t_k}dt_{k-1}\ldots \int_{0}^{t_2}dt_1\\
		&\times\left\lvert \langle
n|e^{-iH_c(t-t_k)}C_{i_k}e^{-iH_c(t_k-t_{n-1})}C_{i_{n-1}}\ldots
C_{i_1}e^{-iH_ct_1}|m\rangle\right\rvert^2P_{ss}(m)\\
		=&\sum_{k=0}^{\infty}
		\sum_{i_1=-1}^{1}\sum_{i_2=-1}^{1}\ldots \sum_{i_k=-1}^{1}
		\int_{0}^{t}dt_k\int_{0}^{t_k}dt_{k-1}\ldots \int_{0}^{t_2}dt_1\\
		&\times\left\lvert \langle m|e^{-iH_c(t-t_k)}C_{-i_1}
		e^{-iH_c(t_k-t_{k-1})}C_{-i_2}\ldots
C_{-i_k}e^{-iH_ct_1}|n\rangle\right\rvert^2P_{ss}(n)
	\end{split}
\end{equation}
where we have used $P_{nm}[\gamma]=P_{mn}[\tilde{\gamma}]$, to arrive
at the last line, and
where the $k=0$ contribution to the sum is to be identified with that
due to the no-jump trajectory.

If we now make a change of summation index $-i_l\to i_{k-l+1}$ we get
	\begin{equation}
	\begin{split}
		P(n,t;m,0)
		=&\sum_{k=0}^{\infty}\sum_{i_1=-1}^{1}\sum_{i_2=-1}^{1}\ldots
\sum_{i_k=-1}^{1}\int_{0}^{t}dt_k\int_{0}^{t_k}dt_{k-1}\ldots
\int_{0}^{t_2}dt_1\\
		&\left\lvert \langle
m|e^{-iH_c(t-t_k)}C_{i_k}e^{-iH_c(t_k-t_{k-1})}C_{i_{k-1}}\ldots
C_{i_1}e^{-iH_ct_1}|n\rangle\right\rvert^2P_{ss}(n)\\
		=&P(m,t;n,0)
	\end{split}
\end{equation}
\end{widetext}
and hence the equality of the two correlation functions,
$g^{(2)}(t)=g_1(t)$ \cite{Cresser94,Herzog94,Cresser96} this equality
coming about since for each trajectory $\gamma$ contributing to one
correlation function, the dual time reversed trajectory
$\tilde{\gamma}$ contributes with equal probability to the other
correlation function.

More succinctly, this result amounts to showing that since, by the
quantum regression theorem we have
\begin{equation}
		\langle a^\dagger(0)a^\dagger(t)
a(t)a(0)\rangle=\text{Tr}_S\left[\mathcal{J}_{-1}e^{\mathcal{L}
t}\mathcal{J}_{-1}\rho_{ss}\right]
\end{equation}
then, on using the above procedure we get
\begin{align}
	\langle a^\dagger(0)a^\dagger(t) a(t)a(0)\rangle
	\propto&\text{Tr}_S\left[\widetilde{\mathcal{J}_{-1}}
	e^{\tilde{\mathcal{L}}t}\widetilde{\mathcal{J}_{-1}}\rho_{ss}\right]\\
	=&\text{Tr}_S\left[\mathcal{J}_1e^{\mathcal{L}t}\mathcal{J}_1\rho_{ss}\right]
\end{align}
the last term here being proportional to $g_1(t)$.

This essentially means that the intensity correlations of the cavity
field are encoded in the correlation properties of the atomic beam
emerging from the cavity.
This same connection between correlation properties of the emergent
atomic beam and the cavity field can also be shown to extend to field
correlations and a homodyne-like experiment performed on the emergent
atomic beam \cite{Cresser06}, i.e., the cavity field spectrum
\cite{Scully03,Casagrande03,Quang05,Vogel05} can also be measured by
studies of atomic beam correlations.

\subsection{More general cases}

If the same approach is adopted in the general case of
$p\ne1,\bar{n}\ne0$, for which the steady state is now given by the
more complex expression Eq.\ (\ref{SteadyStatePhotonNumber}), we find
that the dual of $\mathcal{J}_{-1}$ is no longer readily identifiable
as representing a measurement on the atomic beam, and the above
method will fail. Whether or not a relationship can be found by other
means, or even exists, remains an open question.

If the incident beam is not Poissonian, then the above relationship
also appears not to hold. For instance, if the case of a
super-bunched beam, with $g(t)$ as given by  
Eq.\ (\ref{SuperBunchedg}), for which the steady state is given by
Eq.\ (\ref{SuperBunchedrhoss}), the dual relationship between the
jump operators $C_1$ and $C_{-1}$ ceases to hold, so the above
procedure for constructing time-reversed quantum trajectories will
break down.

\section{A fluctuation relation for the micromaser\label{MicroFT}}

A detailed quantum fluctuation theorem involves relating the
probability $P[\gamma]$ to observe a quantum trajectory $\gamma$ in
the forward time direction to the probability
$\tilde{P}[\tilde{\gamma}]$ of its time-reversed conjugate
$\tilde{\gamma}$ \cite{Horowitz12}. Such a theorem provides a measure
of the irreversibility of the dynamics for a given trajectory. But in
contrast to the previous result, where the notion of a time reversed
trajectory emerged as a natural part of the calculation, there arises
the matter of defining the time-reversed quantum jump operators in
order to derive a detailed fluctuation theorems, in the form of the
ratio Eq.\ (\ref{FluctuationTheorem}). 

The Crooks approach is not without its difficulties as found for
instance in application to driven quantum systems, or systems for
which there is no steady state fixed point \cite{Barra17}. Problems
of a different nature arise here, traced to the fact that the cavity
field is coupled to two reservoirs. If the dual operators are defined
with respect to the steady state $\rho_{ss}$, no meaningful
fluctuation relation arises. This difficulty has been discussed in
\cite{Manzano15}, and the argument can be made that the time-reversed
jump operators ought to be those that satisfy the Crooks condition
(that the forward and reversed trajectories have the same probability
at steady state) for each jump operator in the presence of its
associated reservoir \emph{only}. Thus, for the jump operators
associated with the coupling of the cavity field to the thermal
reservoir, we will require that the duals to the jump operators
$C_{-1}=\sqrt{\gamma (\bar{n}+1)}a$ and
$D_{1}=\sqrt{\gamma\bar{n}}a^\dagger$ be given by
\begin{equation}
\widetilde{C}_{-1}=\Theta\rho_{ss}^{1/2}C_{-1}^\dagger\rho_{ss}^{-1/2}
\Theta^\dagger\big|_{R=0}=D_1=e^{-\hbar\omega/2kT_c}C_{-1}^\dagger
\end{equation}
and
\begin{equation}
\widetilde{D}_1=\Theta\rho_{ss}^{1/2}D_1^\dagger\rho_{ss}^{-1/2}
\Theta^\dagger\big|_{R=0}=C_{-1}=e^{\hbar\omega/2kT_c}D_1^\dagger.
\end{equation}
where
$\rho_{ss}|_{R=0}=(1-e^{\hbar\omega/kT_c})e^{-N\hbar\omega/kT_c}$ is
cavity field steady state in the absence of an atomic beam reservoir.
For the jump operators associated with an atom passing through the
cavity, $C_1=\sqrt{pR}[\sin(\theta\sqrt{N})/\sqrt{N}]a^\dagger$ and
$D_{-1}=\sqrt{(1-p)R}[\sin(\theta\sqrt{N+1})/\sqrt{N+1}]a$ we find
\begin{equation}
\widetilde{C}_1=\Theta\rho_{ss}^{1/2}C_1^\dagger\rho_{ss}^{-1/2}
\Theta^{-1}\big|_{\gamma=0}=D_{-1}=e^{\hbar\omega/2kT_a}C_1^\dagger
\end{equation}
and
\begin{equation}
\widetilde{D}_{-1}=\Theta\rho_{ss}^{1/2}D_{-1}^\dagger\rho_{ss}^{-1/2}
\Theta^{-1}\big|_{\gamma=0}=C_1=e^{-\hbar\omega/2kT_a}D_{-1}^\dagger
\end{equation}
where, from Eq.\ (\ref{SteadyStatePhotonNumber}),
$\rho_{ss}\big|_{\gamma=0}=(1-e^{\hbar\omega/kT_a})e^{-N\hbar\omega/kT_a}$
is the cavity field steady state in the presence of the atomic beam
only. the remaining operators $C_0$ and $D_0$ are unaffected.

With these results at hand it is now straightforward to determine the
ratio of the forward and backward trajectory probabilities. Thus we
wish to compare the two conditional probabilities $P[\gamma_{mn}]$
for the forward trajectory $\gamma_{mn}$ and $P[\tilde{\gamma}_{nm}]$
for the time reversed dual trajectory $\tilde{\gamma}_{nm}$ (i.e.,
excluding boundary terms \cite{Manzano15} depending on the
probabilities of the initial states of the forward and reverse
processes):
\begin{equation}
	\frac{P[\gamma_{mn}]}{P[\tilde{\gamma}_{nm}]}
	=\frac{\left\lvert \langle n|e^{-iH_c(t-t_k)}L_{i_k}\ldots
L_{i_1}e^{-iH_ct_1}|m\rangle\right\rvert^2}{\left\lvert \langle m|
	e^{-iH_c(t-t_k)}\tilde{L}_{i_k}\ldots \tilde{L}_{i_1} 
	e^{-iH_c t_1}|n\rangle\right\rvert^2}
\end{equation}
where $L_i\in\{C_i,D_i\}$. Each pairing in the numerator and
denominator of the operator $C_1$ and its dual $D_{-1}$ will
contribute a factor $e^{-\hbar\omega/kT_a}$, $D_1$ and its dual
$C_{-1}$ a factor $e^{-\hbar\omega/kT_c}$  and so on. As the number
states are eigenstates of $H_c$, and the jump operators map number
states into number states, the remaining factors in the numerator and
denominator cancel exactly and we are left with
\begin{equation}
	P[\tilde{\gamma}_{nm}]
	=\exp[(\Delta E_a(\gamma)/kT_a+\Delta
E_c(\gamma)/kT_c)]P[\gamma_{mn}]
\end{equation}
where $\Delta E_a(\gamma)$ and $\Delta E_c(\gamma)$ are the total
energies gained by the cavity field through cavity reservoir and the
atomic beam induced quantum jumps, respectively in the forward
process. 

This result is independent of the Rabi factors so holds true
irrespective of the strength of the coupling of the field to the
atoms, i.e., it is not a weak system-reservoir interaction result. It
is also independent of the initial and final states, a general result
not specific to the micromaser \cite{Manzano15}, and as such is
dependent solely on the history of quantum jumps, so the subscripts
$m$ and $n$ can be suppressed. Finally, there is no dependence on the
atomic arrival times $t_1,t_2\ldots $. Thus the above result will
remain true even if the arrival times are described by a more general
process (e.g., a renewal process) as it is meaningful to unravel the
system dynamics as an ensemble of quantum trajectories in spite of
the master equation being non-Markovian, as discussed in Section
\ref{UnravelledNonMark}.

The ratios $-\Delta E_c/T_c$ and $-\Delta E_a/T_a$ can be recognized
as the entropy flows $\Delta S_c$ and $\Delta S_a$ from the cavity
reservoir and atomic beam respectively, so that we have
\begin{equation}
	P[\tilde{\gamma}]=e^{-(\Delta S_a(\gamma)+\Delta
S_c(\gamma))/k}P[\gamma]\label{RatioEntropy}
\end{equation}
or equivalently
\begin{equation}
	P[\gamma]=e^{-(\Delta S_a(\tilde{\gamma})+\Delta
S_c(\tilde{\gamma}))/k}P[\tilde{\gamma}]
\end{equation}
an example of a general set of quantum trajectory derived fluctuation
theorems \cite{Horowitz12,Horowitz13,Hekking13,Manzano15,Elouard17}.

It should be made clear that the time reversed trajectories are
explicitly constructed for comparison with the forward trajectories,
but both represent possible physically realisable forward
trajectories. Thus the comparison of the probability of the two
trajectories embodied in Eq.\ (\ref{RatioEntropy}) is a comparison of
two forward trajectories, with one having a reversed sequence of
quantum jumps. So this final result tells us that the trajectory for
which the total entropy change $\Delta S_a+\Delta S_c$ is greater
than zero will be exponentially more likely than its reversed
counterpart, an outcome consistent with the second law. 

\section{Summary and conclusions}

The micromaser, an early example of a collisional or repeated
interaction model of an open quantum system, has been investigated
here with attention paid to an understanding of its properties based
on a time-reversed quantum trajectory analysis. The master equation
for a general impulsive repeated interaction model was rederived in
the general setting of a renewal process describing the interaction
time statistics. The approach developed makes it possible to show the
impact of the underlying assumption of stationary statistics of these
interaction times, edifying  the differences between this approach
and those of other researchers, most recently \cite{Budini09,Vacchini16}, that yield somewhat different master equations, and showing that they are asymptotically in agreement.

This work then provided a background in which the notion of
time-reversed quantum trajectories (TRQT) could be formulated for the
micromaser. The Crooks prescription \cite{Crooks08} for constructing TRQTs is then
shown to arise in a natural fashion (i.e., no put in `by hand') when analysing the relationship between micromaser field correlation properties and those of the
atomic beam, a relationship that breaks down if the incident atomic
beam is not Poissonian.

Attention was then given to using the Crooks approach to define a
class of TRQTs suitable for investigating thermodynamic quantum trajectory
fluctuation relations for the micromaser in the sense of introduced by \cite{Horowitz12}. However, in this case, it is now necessary to \emph{define} what is meant by a time-reversed quantum trajectory. In other words, in contrast to the previous instance, a time-reversed quantum trajectory has to be constructed `by hand'. The fact that the micromaser is a system interacting with two reservoirs also implies that further care be taken in constructing the TRQTs \cite{Manzano15}. The results show that neither the strong coupling of the micromaser field to the atoms, nor non-Poissonian arrival statistics has any impact on the fluctuation relation, which is of the generally expected form.

Further work can focus on the detailed thermodynamic properties of
the micromaser, in particular with respect to the action of the
atomic beam, an the circumstances under which it can be considered a
work reservoir or a thermal reservoir \cite{Strasberg17}. And of
course, a natural extension of this work would be to those cases in
which the atomic beam possesses coherence, or to the phaseonium model
of \cite{Scully03a}.

\section{Acknowledgements}

Thanks go to Steve Barnett who provided much needed feedback on the
topic of time-reversal, and for a reading of an earlier version of
the manuscript. 
\bibliographystyle{apsrev4-1.bst}
\bibliography{/Users/jcresser/Documents/Research/JDCBib.bib}

\appendix

\section{Derivation of non-Markovian master equation\label{MEDerivation}}

Introducing the auxiliary quantity
	\begin{equation}
	\begin{split}
		\sigma(t)=&R^{-1}\langle \langle I(t)\rho_c(t)\rangle\rangle\\
		=&e^{\mathcal{L}_St}\rho(0)+R\int_{0}^{t}
		d\tau e^{\mathcal{L}_S(t-\tau)}g(t-\tau)\mathcal{F}_a \sigma(\tau)
	\end{split}
	\label{sigmaExpression}
\end{equation}
enables Eq.\ (\ref{rhoExpansion}) to be written in terms of $\sigma(t)$ as
\begin{equation}
	\dot{\rho}(t)=\mathcal{L}_S\rho(t)+R\mathcal{F}_a \sigma
	\label{rhoSigma}
\end{equation}
Taking the Laplace transform of Eq.\ (\ref{rhoSigma}) yields, with
$\bar{s}=s-\mathcal{L}_S$
\begin{equation}
	\bar{s}\tilde{\rho}(s)-\rho(0)=R\mathcal{F}_a \tilde{\sigma}(s)
	\label{A1}
\end{equation}
while for Eq.\ (\ref{sigmaExpression}) we get
\begin{equation}
	\tilde{\sigma}(s)
	=\left(1-R\tilde{g}(\bar{s})\mathcal{F}_a\right)^{-1}\bar{s}^{-1}\rho(0).
	\label{A2}
\end{equation}
Combining Eq.\ (\ref{A1}) and (\ref{A2}) by eliminating $\rho(0)$
then gives $\tilde{\sigma}(s)=\tilde{K}(\bar{s})\tilde{\rho}(s)$ where
\begin{equation}
	\tilde{\mathcal{K}}(s)
	=\left(1-(\tilde{g}(s)-s^{-1})R\mathcal{F}_a\right)^{-1}
	\label{K}.
\end{equation}
Substituting this into Eq.\ (\ref{A1}) and inverting the Laplace
transform then gives the required master equation
\begin{equation}
	\frac{d\rho}{dt}=\mathcal{L}_S\rho+R\mathcal{F}_a
	\int_{0}^{t}e^{\mathcal{L}_S(t-\tau)}\mathcal{K}(t-\tau)\rho(\tau)d\tau.
\end{equation}

\section{Expansion of non-Markovian master equation\label{Unravelling}}

The Laplace transform density operator $\tilde{\rho}(s)$ can be
written
\begin{equation}
	\tilde{\rho}(s)=\left(\bar{s}-R\mathcal{F}_a
	\tilde{K}(\bar{s})\right)^{-1}\rho(0)
\end{equation}
which on substituting for $\tilde{\mathcal{K}}$, Eq.\ (\ref{K}) leads
to
\begin{equation}
	\tilde{\rho}(s)=\left(1+\bar{s}^{-1}R\mathcal{F}_a
	\left(1-\tilde{g}(\bar{s})R\mathcal{F}_a\right)^{-1}\right)\bar{s}^{-1}\rho(0).
\end{equation}
From the defining equation for $g(t)$, Eq.\ (\ref{gDefined}), we have
\begin{equation}
	R\tilde{g}(s)=\frac{\tilde{w}(s)}{1-\tilde{w}(s)}
	\label{B3}
\end{equation}
which can be used to reduce the expression for $\tilde{\rho}(s)$ to
\begin{multline}
	\tilde{\rho}(s)=\bar{s}^{-1}
	\left(1-R\frac{1-\tilde{w}(\bar{s})}{\bar{s}}\right)\rho(0)\\
	+\frac{1-\tilde{w}(\bar{s})}{\bar{s}}(1+\mathcal{F}_a )
	\left(1-\tilde{w}(\bar{s})
	(1+\mathcal{F}_a)\right)^{-1}R\frac{1-\tilde{w}(\bar{s})}{\bar{s}}\rho(0).
\end{multline}
Inverting the Laplace transform then gives
\begin{widetext}
	\begin{align}
	\rho(t)=&e^{\mathcal{L}_St}
	\left(1-\int_{0}^{t}d\tau
	R\left(1-\int_{0}^{\tau}d\tau'w(\tau')\right)\right)\rho(0)\nonumber\\
	&+\sum_{n=1}^{\infty}\int_{0}^{t}dt_n\int_{0}^{t_n}dt_{n-1}\ldots
	\int_{0}^{t_2}dt_1\nonumber\\
	&\times\left(1-\int_{0}^{t-t_n}d\tau
	w(\tau)\right)w(t_n-t_{n-1})\ldots
	w(t_2-t_1)R\left(1-\int_{0}^{t_1}d\tau'w(\tau')\right)\nonumber\\ 
	&\times e^{\mathcal{L}_S(t-t_n)}(1+\mathcal{F}_a)
	e^{\mathcal{L}_S(t_n-t_{n-1})}(1+\mathcal{F}_a)\ldots
	(1+\mathcal{F}_a)e^{\mathcal{L}_St_1}\rho(0).
\end{align}
\end{widetext}
The interpretation of this expansion is given in Section \ref{ReProMaEq}.
\end{document}